\documentstyle[aps]{revtex}

\begin{document}
\title{Potential Models for Radiative Rare $B$ Decays}
\author{Saeed Ahmad\thanks{%
email: sahmad76@yahoo.com}}
\address{{\em Department of Physics and National Centre for Physics, Quaid-i-Azam}\\
University, Islamabad 45320, Pakistan}
\author{Riazuddin\thanks{%
email: ncp@comsats.net.pk}}
\address{{\em National Centre for Physics, Quaid-i-Azam University, Islamabad 45320,}%
\\
Pakistan}
\date{NCP-QAU/19-1000}
\maketitle

\begin{abstract}
We compute the branching ratios for the radiative rare decays of $B$ into $K$%
-Meson states and compare them to the experimentally determined branching
ratio for inclusive decay $b\rightarrow s\gamma \ $using non relativistic
quark model, and form factor definitions consistent with HQET covariant
trace formalism. Such calculations necessarily involve a potential model. In
order to test the sensitivity of calculations to potential models we have
used three different potentials, namely linear potential, screening
confining potential and heavy quark potential as it stands in QCD.We find
the branching ratios relative to the inclusive $b\rightarrow s\gamma $ decay
to be $(16.07\pm 5.2)\%$ for $B\rightarrow K^{\star }(892)\gamma $ and $%
(7.25\pm 3.2)\%$ for $B\rightarrow K_2^{\star }(1430)\gamma $ for linear
potential. In the case of the screening confining potential these values are 
$(19.75\pm 5.3)\%$ and $(4.74\pm 1.2)\%$ while those for the heavy quark
potential are $(11.18\pm 4.6)\%$ and $(5.09\pm 2.7)\%$ respectively. All
these values are consistent with the corresponding present CLEO experimental
values: $(16.25\pm 1.21)\%$ and $(5.93\pm 0.46)\%.$\\\\PACS number(s):
13.40.Hq, 12.37.Pn, 12.39.Hg
\end{abstract}

\section{Introduction}

The flavour changing weak decays of mesons have always been a rich source of
information about basic interactions in particle physics. In particular,
radiative $B$ decays $B\rightarrow K^{**}\gamma $ $(K^{**}\sim K^{*}(892),$ $%
K^{*}(1430)$ etc.) received intensive theoretical studies. The presence of
heavy $b$ quark permits the use of Heavy Quark Effective Theory (HQET) in
evaluating the relevant hadronic matrix elements where the relevance of the
use of a potential model comes in. One purpose of our paper is to test the
sensitivity of the branching ratios for $B\rightarrow K^{**}\gamma $ decays
relative to the inclusive $b\rightarrow s\gamma $ decay rate to potential
models. Among the radiative processes $B\rightarrow X_s\gamma ,B\rightarrow
K^{*}(892)\gamma $ and $B\rightarrow K_2^{*}(1430)\gamma $ exclusive
branching ratios have been measured experimentally\cite{1}:

\begin{equation}
{\cal B}(B\rightarrow K^{*}(892)\gamma )=(4.55\pm 0.34)\times 10^{-5}
\end{equation}

\begin{equation}
{\cal B}(B\rightarrow K_2^{*}(1430)\gamma )=(1.66\pm 0.13)\times 10^{-5}
\end{equation}
and so has been the inclusive rate\cite{2}

\begin{equation}
{\cal B}(B\rightarrow X_{s}\gamma )=(2.32\pm 0.57\pm 0.35)\times 10^{-5}
\end{equation}

Several methods have been employed to predict $B\rightarrow K^{*}\gamma $
decay rate: HQET\cite{3,4}, QCD sum rules\cite{5}-\cite{10}, quark models%
\cite{11}-\cite{21}, bound state resonances\cite{22} and Lattice QCD \cite
{23}-\cite{26}. In this paper we follow the approach of \cite{3,4} in which
both $b$ and $s$ quarks are considered heavy. Although the $s$-quark is not
particularly heavy and very substantial corrections to the Isgur-Wise
functions are to be anticipated, particularly at the symmetry point, yet it
has been found that the use of heavy quark symmetry for $s$-quark has not
been too bad \cite{4}, particularly in connection with prediction of decay
rates for $D\rightarrow K^{(*)}\ell \nu _\ell $,$D\rightarrow K\ell \nu
_\ell $ from $B\rightarrow D^{(*)}\ell \nu $ \cite{3}. It is, therefore, not
an unreasonable hope that the static limit may provide results of a
comparable accuracy also for the radiative rare $B$-decays. In fact the
agreement obtained for $B\rightarrow K^{*}\gamma $ seems to support this
hope. In any case the rates of $K^{**}$ states relative to $B\rightarrow
K^{*}\gamma ,$ should not too much depend on $\frac 1{m_s}$ corrections. In
the heavy quark limit, the long distance effects are contained within
unknown form factors whose precise definition consistent with the covariant
trace formulasim \cite{27}- \cite{30} has now been clarified.We use the same
non relativistic quark model for the wave functions of the light degree of
freedom (LDF) as was used in \cite{4} but employ the numeical approach of 
\cite{31}.We use three different potentials, linear potential $``V=\frac{%
-4\alpha _s}{3r}+br+c"$, screening confiningpotential $``V=\left( \frac{%
-4\alpha _s}{3r}+br\right) \frac{1-e^{-\mu r}}{\mu r}"\cite{19}$ and heavy
quark pothential $``V=br-\frac{8C_F}ru(r)"$\cite{32}. Our results for linear
potential are $(16.07\pm 5.2)\%$ for $B\rightarrow K^{*}(892)\gamma $and $%
(7.25\pm 3.2)\%$ for $B\rightarrow K_2^{*}(1430)\gamma $ while those for
screening confining potential are $(19.75\pm 5.2)\%$and $(4.74\pm 1.2)\%$
and for heavy quark potential they are $(11.18\pm 4.6)\%$and $(5.09\pm
2.7)\% $ respectively. They are in good agreement with the recent
experimental measurements made by CLEO \cite{1} and at the present precision
of both theoretical and experimental values, one cannot distinguish between
the above potentials. However, when the data for other decays of $B$ into
some of the other $K$-meson states become available, it should be possible
to make such a distinction-[see Table 1].

In section II we present the theoretical framework for $B\rightarrow
K^{**}\gamma $ decays. ISGW Model is described in section III which also
contains the procedure for finding wave function of LDF using different
potential models. The obtained results are summarized in the Conclusion.

\section{Theoretical Framework}

For $B\rightarrow K^{**}\gamma $ decays the effective hamiltonian is well
known \cite{31} \cite{33}-\cite{34}:

\begin{equation}
H_{eff}=-\frac{4G_F}{\sqrt{2}}V_{tb}V_{ts}^{*}C_7(m_b){\cal O}_7(m_b)
\end{equation}
where 
\begin{equation}
{\cal O}_7{\cal =}\frac e{32\pi ^2}F_{\mu \nu }\left[ m_b\stackrel{\_}{s}%
\sigma ^{\mu \nu }(1+\gamma _5)b+m_s\stackrel{\_}{s}\sigma ^{\mu \nu
}(1-\gamma _5)b\right]
\end{equation}
Matrix elements of bilinear currents of two heavy quarks $(J(q)=\stackrel{\_%
}{Q}TQ)$ are most conveniently evaluated within the framework of covarient
trace formulasim. Denoting $\omega =v^{\prime }.v,$ where $v$ and $v^{\prime
}$ are the four velocities of the two mesons, we have

\[
\langle \Psi ^{\prime }(v^{\prime })\left| J(q)\right| \Psi (v)\rangle =Tr[%
\stackrel{\_}{M}(v^{\prime })\Gamma M(v)]{\cal M}(\omega ) 
\]
where $M^{\prime }$ and $M$ denote matrices describing states $\Psi ^{\prime
}(v^{\prime })$ and $\Psi (v)$, $\stackrel{\_}{M}$=$\gamma ^0M^{\dagger
}\gamma ^0,$ and ${\cal M}(\omega )$ represents the LDF. $M$, $M^{\prime }$
and ${\cal M}(\omega )$ can be found in \cite{4,30} and using Eq.(5) we can
write:

\begin{equation}
\langle K^{**}\gamma \left| {\cal O}_7(m_b)\right| B\rangle =\frac e{16\pi ^2%
}\eta _\mu q_\nu Tr[\stackrel{\_}{M^{\prime }}(v^{\prime })\Omega ^{\mu \nu
}M(v)]{\cal M}(\omega )
\end{equation}
where the factor $q_\nu =m_Bv_\nu -m_{K^{**}}v_\nu ^{\prime }$ comes from
the derivative in the field strength F$_{\mu \nu }$ of Eq.(4), $\eta _\mu $
is the photon polarization vector and

\begin{equation}
\Omega ^{\mu \nu }=m_B\sigma ^{\mu \nu }(1+\gamma _5)+m_{K^{**}}\sigma ^{\mu
\nu }(1-\gamma _5)
\end{equation}
Using the mass shell condition of the photon $(q^2=0)$ and polarization sums
for spin-1 and spin-2 particles, the decay rates are calculated in \cite
{4,31}:

\begin{equation}
\Gamma (B\rightarrow K^{*}(892)\gamma )=\Omega \left| \xi _{C}(\omega
)\right| ^{2}\frac{1}{y}[(1-y)^{3}(1+y)^{5}(1+y^{2})]
\end{equation}

\begin{equation}
\Gamma (B\rightarrow K_{1}(1270)\gamma )=\Omega \left| \xi _{E}(\omega
)\right| ^{2}\frac{1}{y}[(1-y)^{5}(1+y)^{3}(1+y^{2})]
\end{equation}

\begin{equation}
\Gamma (B\rightarrow K_{1}(1400)\gamma )=\Omega \left| \xi _{F}(\omega
)\right| ^{2}\frac{1}{24y^{3}}[(1-y)^{5}(1+y)^{7}(1+y^{2})]
\end{equation}
\begin{equation}
\Gamma (B\rightarrow K_{2}^{*}(1430)\gamma )=\Omega \left| \xi _{F}(\omega
)\right| ^{2}\frac{1}{8y^{3}}[(1-y)^{5}(1+y)^{7}(1+y^{2})]
\end{equation}

\begin{equation}
\Gamma (B\rightarrow K^{*}(1680)\gamma )=\Omega \left| \xi _{G}(\omega
)\right| ^{2}\frac{1}{24y}[(1-y)^{7}(1+y)^{5}(1+y^{2})]
\end{equation}

\begin{equation}
\Gamma (B\rightarrow K_{2}(1580)\gamma )=\Omega \left| \xi _{G}(\omega
)\right| ^{2}\frac{1}{8y^{3}}[(1-y)^{7}(1+y)^{5}(1+y^{2})]
\end{equation}

\begin{equation}
\Gamma (B\rightarrow K^{*}(1410)\gamma )=\Omega \left| \xi _{C_{2}}(\omega
)\right| ^{2}\frac{1}{y}[(1-y)^{3}(1+y)^{5}(1+y^{2})]
\end{equation}

\begin{equation}
\Gamma (B\rightarrow K_1(1650)\gamma )=\Omega \left| \xi _{E_2}(\omega
)\right| ^2\frac 1y[(1-y)^5(1+y)^3(1+y^2)]
\end{equation}
where

\begin{equation}
y=\frac{m_{K^{**}}}{m_{B}}
\end{equation}

\begin{equation}
\Omega =\frac \alpha {128\pi ^4}G_F^2m_B^5\left| V_{tb}\right| ^2\left|
V_{ts}\right| ^2\left| C_7(m_B)\right| ^2
\end{equation}
and the argument of the Isgur-Wise (IW) function is fixed by the mass shell
condition of the photon $(q^2=0),$

\begin{equation}
\omega =\frac{1+y^{2}}{2y}
\end{equation}

\section{Potential Models for the Isgur-Wise Functions}

Following \cite{31} for the evaluation of IW form factors needed for the
decay rates we assume that we can describe heavy-light mesons using some
non-relativistic potential models; the rest frame LDF wave function can then
be written as

\begin{equation}
\phi _{j\lambda j}^{(\alpha L)}(x)=\sum_{m_L,m_s}R_{\alpha
L}(r)Y_{Lm_L}(\Omega )\chi _{m_s}\langle L,m_L;\frac 12,m_s\mid j,\lambda
_j;L,\frac 12\rangle
\end{equation}
where $\chi _s$ represent the rest frame spinors normalized to one, $\chi
_{m_s^{^{\prime }}}^{\dagger }\chi _{m_s}=\delta _{m_s^{^{\prime }},m_s}$
and $\alpha $ represents all other quantum numbers. In \cite{31} following
expressions for the form factors are obtained: 
\begin{equation}
\xi _C(\omega )=\frac 2{\omega +1}\langle j_0(ar)\rangle _{00},\qquad \qquad
\qquad \qquad 0_{\frac 12}^{-}\rightarrow (0_{\frac 12}^{-},1_{\frac 12}^{-})
\end{equation}

\begin{equation}
\xi _{F}(\omega )=\frac{2}{\sqrt{\omega ^{2}-1}}\langle j_{1}(ar)\rangle
_{10},\qquad \qquad \quad \quad \quad 0_{\frac{1}{2}}^{-}\rightarrow (0_{%
\frac{1}{2}}^{+},1_{\frac{1}{2}}^{+})
\end{equation}

\begin{equation}
\xi _{F}(\omega )=\sqrt{\frac{3}{\omega ^{2}-1}}\frac{2}{\omega +1}\langle
j_{1}(ar)\rangle _{10},\qquad \qquad 0_{\frac{1}{2}}^{-}\rightarrow (1_{%
\frac{3}{2}}^{+},2_{\frac{3}{2}}^{+})
\end{equation}

\begin{equation}
\xi _G(\omega )=\frac{2\sqrt{3}}{\omega ^2-1}\langle j_2(ar)\rangle
_{20},\qquad \qquad \qquad \quad \quad 0_{\frac 12}^{-}\rightarrow (1_{\frac 
32}^{-},2_{\frac 32}^{-})
\end{equation}
where denoting the energy of LDF as $E_q:$

\begin{equation}
a=(E_{q}+E_{\stackrel{\_}{q}^{^{\prime }}})\sqrt{\frac{\omega -1}{\omega +1}}
\end{equation}
and

\begin{equation}
\langle F(ar)\rangle _{L^{^{\prime }}L}^{\alpha ^{^{\prime }}\alpha }=\int
r^{2}drR_{\alpha ^{^{\prime }}L^{^{\prime }}}^{*}(r)R_{\alpha L}(r)F(ar)
\end{equation}
To find the form factors we use the method of \cite{31} i.e. solve the
Schrodinger Equation numerically.We use three different potentials

Linear potential: 
\begin{equation}
V=\frac{-4\alpha _s}{3r}+br+c,
\end{equation}

Screening confining potential \cite{19}: 
\begin{equation}
V=\left( \frac{-4\alpha _s}{3r}+br\right) \frac{1-e^{-\mu r}}{\mu r},
\end{equation}

and Heavy quark potential \cite{32}: 
\begin{equation}
V=br-\frac{8C_F}ru(r)
\end{equation}
Following \cite{31}, we fix $b=0.18$ $GeV^2$ and vary $\alpha _s$ and $c$
for a given value of $m_{u,d}$(in the range $0.30-0.35GeV$) and $m_s$(in the
range $0.5-0.6GeV$), untill a good description of the spin averaged spectra
of $K$-meson states is obtained. Following this procedure, our $\alpha _s$
ranges from $0.37$ to $0.48$, while $c$ takes values from $-0.83GeV$ to $%
-0.90GeV$. These parameters are in good agreement with the original ISGW
values \cite{36}($\alpha _s=0.5$ and $c=-0.84GeV$ for $m_{u,d}=0.33$ $GeV$
and $m_s=0.55GeV$). For the screening confining potential: $\sigma =0.18\pm
0.02$ $GeV^2$ and $\mu ^{-1}=0.8\pm 0.2$ $fm$, while for the heavy quark
potential \cite{32} $C_F=\frac{N_c^2-1}{2N_c}$ and [$a(q^2)$ is defined in 
\cite{32}]

\begin{equation}
u(r)=\int_0^\infty \frac{dq}q(a(q^2)-\frac k{q^2})\sin (q.r)
\end{equation}
which is calculated numerically at $r\geq 0.01$ fm and represented in the
MATHEMATICA file in the format of note book at the site http://www.ihep.su/%
\symbol{126}kiselev/Potential.nb. The short distance behaviour of the
potential is purely peturbative, so that at $r\leq 0.01$ fm we can put

\begin{equation}
V(r)=-C_F\frac{\stackrel{\_}{\alpha _v}(1/r^2)}r
\end{equation}
where the value of the running coupling constant $\stackrel{\_}{\alpha _v}%
(1/r^2)$ at $r_s=0.01fm$ is $\stackrel{\_}{\alpha _v}(1/r_s^2)=0.22213.$\\As
in \cite{4} the definition of the LDF energy for a $K^{**}$ meson, proposed
to account for the fact that $s$ mesons are not particulaly heavy, is

\begin{equation}
E_{\stackrel{\_}{q}}=\frac{m_{K^{**}}\times m_{u,d}}{m_s+m_{u,d}}
\end{equation}
Another definition which is consistent with heavy quark symmetry is 
\begin{equation}
E_{\stackrel{\_}{q}}=m_{K^{**}}-m_s
\end{equation}
These two definitions are not equivalent in the heavy quark limit, so we
have done all calculations employing both of these two definitions and at
the end we have quoted the broadest possible range of the results obtained.
Finally, $E_{\stackrel{\_}{q}}$ for the $B$ meson has been taken to be the
same as for the $K^{*}$ meson, consistent with heavy quark symmetry. It
turns out that this is actually a very reasonable assumption. To find the
size of a meson, which we need for the evaluation of the integral in
Eq.(25), we investigate for the asymptotic behavior of Schrodinger Equation
for a particular potential model.\ As an example we display the asymptotic
behavior of LDF wave function for linear potential model(for $\ell =1$) in
Fig 1.

\section{Conclusion}

In table $1$ we present our results for the ratio $R=\frac{\Gamma
(B\rightarrow K^{\star \star }\gamma )}{\Gamma (B\rightarrow X_s\gamma )}$
for various $K$ meson states; the inclusive branching ratio is usually taken
to be QCD improved quark decay rate for $B$ which can be written as \cite
{29,33}

\begin{equation}
\Gamma (B\rightarrow X_s\gamma )=4\Omega (1-\frac{m_s^2}{m_b^2})^3(1+\frac{%
m_s^2}{m_b^2})
\end{equation}
giving the prediction for $BR(b\rightarrow s\gamma )$ to be $(2.8\pm
0.8)\times 10^{-4}$ $\cite{31}$, where the uncertainity is due to the choice
of the QCD scale.

We find that the radiative decays of $B$ into $K$ meson states saturate $30\%
$ to $50\%$ the inclusive decay rate . We cannot reach more quantative
conclusions due to errors involved in theoretical estimates, due to which
also we cannot at present distinguish between the potential models used.
However, even with the present accuracy it should be possible to make such a
distinction when data for the other $K$-states listed in Table $1$ become
available.

\begin{table}[tbp]
\caption{Branching ratios [\%]}
\begin{tabular}{||c|c|c|c|c|c|c||}
Meson & $J^P$ & LinearPot. & Scr.Pot. & Heavy Quark & Ref\cite{31} & 
Experimental value \\ \hline
$K$ & 0$^{-}$ &  & Forbidden &  &  & Ref\cite{1} \\ \hline
$K^{\star }$ & 1$^{-}$ & 16.07$\pm 5.2$ & 19.75$\pm 5.3$ & 11.18$\pm 4.6$ & 
16.82$\pm 6.4$ & 16.25$\pm 1.21$ \\ \hline
$K^{\star }$ & 0$^{+}$ &  & Forbidden &  &  &  \\ \hline
$K_1$ & 1$^{+}$ & 4.68$\pm 0.8$ & 8.33$\pm 2.4$ & 1.63$\pm 0.8$ & 4.28$\pm
1.6$ &  \\ \hline
$K_1$ & 1$^{+}$ & 2.43$\pm 0.8$ & 3.83$\pm 0.8$ & 5.49$\pm 0.8$ & 2.07$\pm
0.9$ &  \\ \hline
$K_2^{\star }$ & 2$^{+}$ & 7.25$\pm 3.2$ & 4.74$\pm 1.2$ & 5.09$\pm 2.7$ & 
6.18$\pm 2.9$ & 5.93$\pm 0.46$ \\ \hline
$K^{\star }$ & 1$^{-}$ & 0.5$\pm 0.2$ & 3.79$\pm 1.1$ & 1.37$\pm 0.4$ & 0.54$%
\pm 0.2$ &  \\ \hline
$K_2$ & 2$^{-}$ & 1.89$\pm 0.6$ & 0.74$\pm 0.2$ & 1.8$\pm 0.4$ & 1.64$\pm
0.4 $ &  \\ \hline
$K$ & 0$^{-}$ &  & Forbidden &  &  &  \\ \hline
$K^{\star }$ & 1$^{-}$ & 5.5$\pm 0.9$ & 6.41$\pm 1.2$ & 3.11$\pm 0.6$ & 4.07$%
\pm 0.6$ &  \\ \hline
$K_0^{\star }$ & 0$^{+}$ &  & Forbidden &  &  &  \\ \hline
$K_1$ & 1$^{+}$ & 1.78$\pm 0.7$ & 3.14$\pm 1.1$ & 2.04$\pm 0.9$ & 1.68$\pm
0.6$ &  \\ \hline
&  & 40.11 & 49.68 & 30.25 & 37.28 & 
\end{tabular}
\end{table}
\begin{figure}[tbp]
\caption{Low r and high r (asymptotic) behavior of LDF wavefunction}
\end{figure}

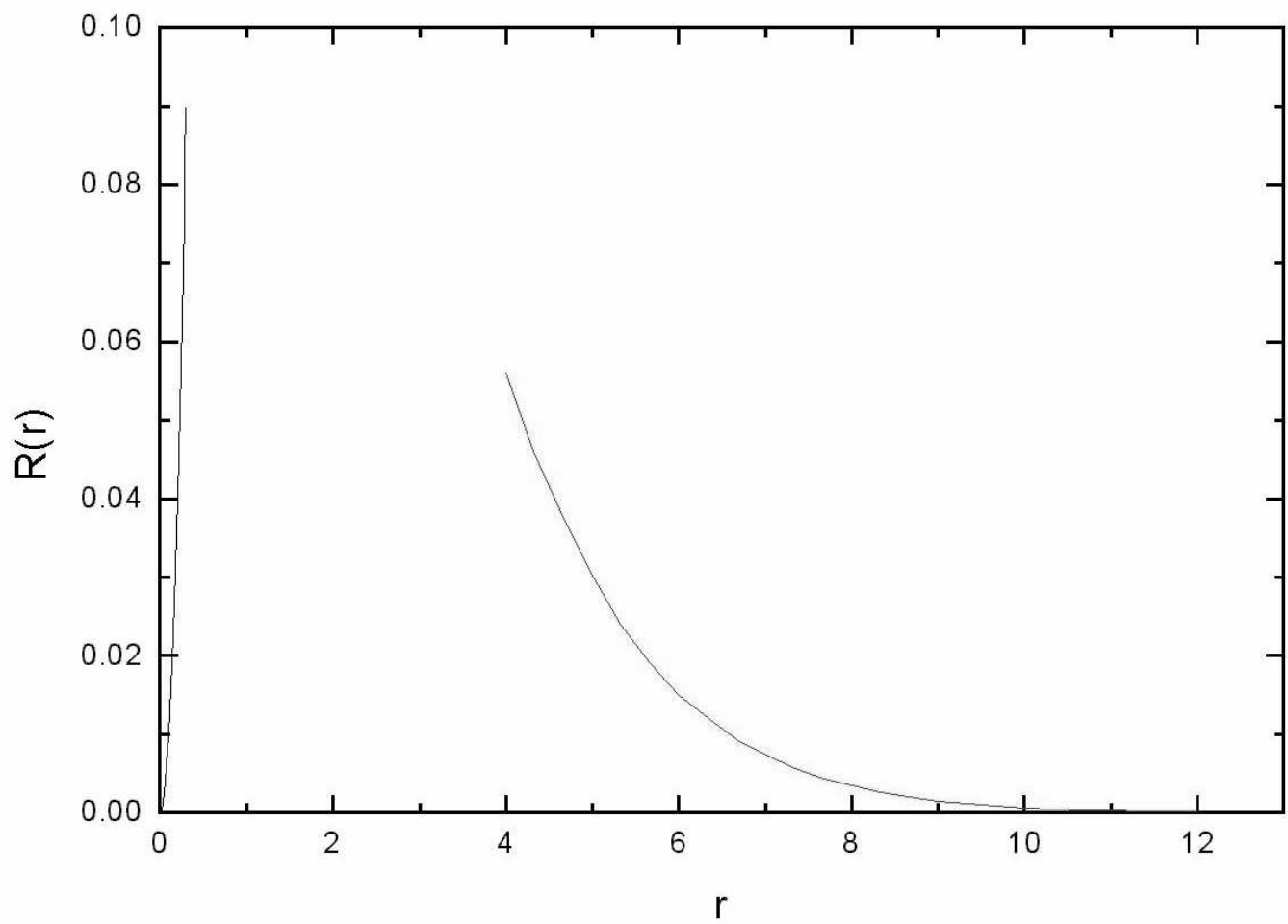
\end{document}